\documentclass[nofootinbib,aps,12pt]{revtex4-1}
\usepackage{amsmath}
\usepackage{amssymb}
\usepackage[left=1cm,right=1cm]{geometry}
\usepackage{feynmp}
\usepackage{slashed}
\usepackage{color}
\usepackage{graphicx}
\usepackage{makecell}
\usepackage{multirow}
\newcommand{\GeV}{\text{GeV}}
\newcommand{\TeV}{\text{TeV}}

\newcommand{\eV}{\text{eV}}

\newcommand{\invfb}{\text{fb}^{-1}}

\begin{document}
\title{\LARGE Dark matter and LHC phenomenology \\
of a scale-invariant scotogenic model}
\bigskip
\author{Chao Guo~$^{1}$}
\email{chaog@mail.nankai.edu.cn}
\author{Shu-Yuan Guo~$^{1,2}$}
\email{shyuanguo@mail.nankai.edu.cn}
\author{Yi Liao~$^{1,2}$}
\email{liaoy@nankai.edu.cn}
\affiliation{
$^{1}$~School of Physics, Nankai University, Tianjin 300071, China
\\
$^2$ Center for High Energy Physics, Peking University, Beijing 100871, China
}
\date{\today}

\begin{abstract}
We study the phenomenology of a model that addresses the neutrino mass, dark matter, and generation of the electroweak scale in a single framework. Electroweak symmetry breaking is realized via the Coleman-Weinberg mechanism in a classically scale invariant theory, while the neutrino mass is generated radiatively through interactions with dark matter in a typically scotogenic manner. The model introduces a scalar triplet and singlet and a vector-like fermion doublet that carry an odd parity of $Z_2$, and an even parity scalar singlet that helps preserve classical scale invariance. We sample over the parameter space by taking into account various experimental constraints from the dark matter relic density and direct detection, direct scalar searches, neutrino mass, and charged lepton flavor violating decays. We then examine by detailed simulations possible signatures at the LHC to find some benchmark points of the free parameters. We find that the future high-luminosity LHC will have a significant potential in detecting new physics signals in the dilepton channel.
\end{abstract}

\maketitle

\section{Introduction}

The existence of tiny neutrino mass and dark matter (DM) provides two pieces of evidence beyond the standard model (SM). Moreover, the SM is afflicted with some theoretical flaws such as the naturalness of the electroweak scale around hundreds of GeV. Hence, it would be interesting to investigate whether it is possible to address these issues within a single framework.

Electroweak symmetry breakdown could occur via the Coleman-Weinberg mechanism~\cite{Coleman:1973jx}, in which the electroweak scale is induced in a scale-invariant theory through radiative effects instead of being put in by hand through a quadratic term of the wrong sign. In recent years, there have been multiple attempts to incorporate DM in the setting of the neutrino mass model, at one-loop~\cite{one loop}, two-loop~\cite{two loops} and higher-loop levels~\cite{three loops}. The idea is to induce a tiny radiative neutrino mass through interactions with new heavy particles that are protected by a global symmetry, so that the lightest of the new particles could serve as a DM candidate. In the minimal model, for instance, a scalar doublet and fermion singlets are introduced. This was generalized in Ref.~\cite{Law:2013saa} by restricting the SM representations of new particles to be no larger than the adjoint representation. Recently, the authors of Ref.~\cite{Ahriche:2016cio} proposed to marry the minimal scotogenic model with the idea of scale invariance by assuming a new scalar singlet. This scalar singlet plays an important role in triggering the radiative breakdown of scale invariance while the lightest fermion singlet serves as DM. The parameter space was found to be severely constrained by direct detection experiments of DM, and viable regions of parameters exist for a DM mass either smaller than $\sim 10~\GeV$ or larger than $\sim 200~\GeV$. Other usages of scale invariance in the context of neutrino mass and dark matter and related extensions of the SM are provided in Refs.~\cite{SInu,SIDM,SInuDM} and the references therein.

From a practical point of view scale invariance reduces the number of free parameters in a model with multiple scalars. In this work, we incorporate scale invariance into a non-minimal scotogenic model suggested in Ref.~\cite{Law:2013saa}. We add a new scalar singlet $\phi$ to preserve scale invariance on top of the new fields already introduced, i.e., a scalar triplet $\Delta$ and singlet $\eta$ plus a vector-like fermion doublet $F$, all of which are protected by a $Z_2$ symmetry. Such a model appeared as one of the various realizations of scale invariant scotogenic models in Ref.~\cite{Ahriche:2016ixu}. As in the general case with multiple scalars~\cite{Gildener:1976ih}, the SM Higgs doublet and the singlet $\phi$ can develop vacuum expectation values (VEVs) radiatively, thus spontaneously breaking scale invariance and electroweak symmetry and generating masses for all particles including neutrinos and DM in particular. The lighter of the singlet $\eta$ and the neutral component of the triplet $\Delta$ could serve as DM. We study various constraints on the new particles and interactions, and investigate the feasibility of detecting the triplet scalars and vector-like fermions through multi-lepton signatures at the LHC.

The paper is organized as follows. In section~\ref{sec:MD}, we introduce the model and discuss radiative breakdown of scale invariance and electroweak symmetry. The radiative neutrino mass and constraints from lepton-flavor-violating (LFV) processes are briefly addressed. In section~\ref{sec:DM}, we study the parameter space that survives the most stringent constraints coming from relic density and direct detection of DM. The feasibility to detect new particles at LHC is simulated in section~\ref{sec:LHC} using the multi-lepton signatures. We summarize our main results in section~\ref{sec:conclusion}.

\section{Model and Constraints}\label{sec:MD}

The scotogenic model, on which our work is based was proposed in Ref.~\cite{Law:2013saa}. It extends the content of the SM fields by an $SU(2)_L$ triplet $\Delta$, a scalar singlet $\eta$ and a vector-like fermion doublet $F$. An exact $Z_2$ parity is assigned to these new fields to stabilize the lightest neutral particle as DM. We introduce a new scalar singlet $\phi$ that helps preserve scale invariance at the classical level, but spontaneously breaks it at the quantum level. The quantum numbers of these fields together with the SM Higgs doublet $\Phi$ and left-handed lepton doublet $L_L$ are collected in table~\ref{tab:ptcl_contents}.

\begin{table}[htbp]
 \caption{\label{tab:ptcl_contents}Relevant fields and their quantum numbers under the SM group and a new $Z_2$ group.}
 \setlength{\tabcolsep}{4mm}{
 \begin{tabular}{c|cc|cccc}
  \hline
  & $L_L$ & $\Phi$ & $F$ & $\eta$ & $\Delta$ & $\phi$ \\
  \hline
 $SU(2)_L$  & 2 & 2 & 2 & 1 & 3 & 1\\
 $U(1)_Y$ & $\displaystyle-\frac{1}{2}$ & $\displaystyle\frac{1}{2}$ & $\displaystyle-\frac{1}{2}$ & 0 & 1 & 0\\
 $Z_2$ & $+$ & $+$ & $-$ & $-$ & $-$ & $+$ \\
  \hline
 \end{tabular}
 }
\end{table}

The general Yukawa interactions involving the new vector-like fermion are given by
\begin{equation}
	-\mathcal{L}_{\text{Yuk.}}\supset y_\eta \overline{F_R} L_L \eta + y_\Delta\overline{L_L^c} \epsilon\Delta F_L + y_\phi\overline{F_L}  F_R \phi+\text{h.c.},
\end{equation}
where $L_L^c$ is the charge conjugation of $L_L$ and $\epsilon$ is the antisymmetric tensor so that $\epsilon L_L^c$ transforms as an $SU(2)_L$ doublet. The Yukawa couplings $y_{\eta,\Delta,\phi}$ are complex matrices with respect to the lepton flavor and new fermion indices. The $F$ and $\Delta$ fields are cast in the form
\begin{equation}
F=\begin{pmatrix}
N \\
E^-
\end{pmatrix},\quad
\Delta=
	\begin{pmatrix}
	\Delta^+/\sqrt{2} & \Delta^{++} \\
	\Delta^0 & -\Delta^+/\sqrt{2}
	\end{pmatrix}.
\end{equation}
A mass term for $F$ is forbidden by scale invariance, and it does not mix with the SM leptons due to the $Z_2$ symmetry. Similarly, $\Phi$ and $\phi$ on the one side do not mix with $\eta$ and $\Delta$ on the other side. Since $\eta$ is pure real, $F$ would have to carry one unit of the lepton number to keep it conserved in the $y_\eta$ term. This would in turn require $\Delta$ to have two negative units of lepton number in the $y_\Delta$ term. However, then a term linear in $\Delta$ (the $\lambda_{\eta\Delta\Phi}$ term) in the scalar potential would still break the lepton number. As shown in the following, the Majorana neutrino mass generated at one loop is indeed proportional to all of the couplings mentioned above.

The complete scale and $Z_2$ invariant scalar potential is generally given as:
\begin{eqnarray}
V&=& \lambda_\Phi\left|\Phi\right|^4
+\lambda_{\Delta 1}\left[\text{Tr}
\left(\Delta^\dagger\Delta\right)\right]^2
+\lambda_{\Delta 2}\text{Tr}
\left(\Delta^\dagger\Delta\Delta^\dagger\Delta\right)
+\frac{1}{4}\lambda_\eta\eta^4 +\frac{1}{4}\lambda_\phi\phi^4
\notag\\
&&+\lambda_{\Phi\Delta 1}\left|\Phi\right|^2
\text{Tr}\left(\Delta^\dagger\Delta\right)
+\lambda_{\Phi\Delta 2}\Phi^\dagger\Delta^\dagger\Delta\Phi +\frac{1}{2}\lambda_{\Phi\eta}\left|\Phi\right|^2\eta^2
+\frac{1}{2}\lambda_{\Phi\phi}\left|\Phi\right|^2\phi^2
\notag\\
& &+\frac{1}{2}\lambda_{\eta\Delta}\eta^2 \text{Tr}\left(\Delta^\dagger\Delta\right)
+\frac{1}{2}\lambda_{\phi\Delta}\phi^2\text{Tr}
\left(\Delta^\dagger\Delta\right)
+\frac{1}{4}\lambda_{\eta\phi}\eta^2\phi^2
+\lambda_{\eta\Delta\Phi}\eta\left(\tilde{\Phi}^\dagger \Delta^\dagger\Phi+\text{h.c.}\right),
\end{eqnarray}
with the trace taken in weak isospin space. We follow the approach in Ref.~\cite{Gildener:1976ih} that generalizes the study of Ref.~\cite{Coleman:1973jx} to the case with multiple scalars. Only the $Z_2$ even scalar fields $\Phi$ and $\phi$ can develop a nonvanishing VEV. Requiring that the VEVs $\langle\Phi\rangle=v/\sqrt{2}$ and $\langle\phi\rangle=u$ occur in the flat direction, where the above tree level potential vanishes, one must have $\lambda_{\Phi\phi}^2=4\lambda_\Phi\lambda_\phi$ with $\lambda_\Phi,~\lambda_\phi>0>\lambda_{\Phi\phi}$. The ratio of the VEVs is given by
\begin{equation}	\frac{v^2}{u^2}=-\frac{\lambda_{\Phi\phi}}{2\lambda_\Phi}
=-\frac{2\lambda_\phi}{\lambda_{\Phi\phi}},
\label{eq:vev}
\end{equation}
while their absolute values are determined by higher order terms in the effective potential. Note that $u$ does not contribute to the masses of the weak gauge bosons, since $\phi$ is a neutral singlet, and $v\approx 246~\GeV$.

\begin{figure}
	\centering \includegraphics[width=0.45\textwidth]{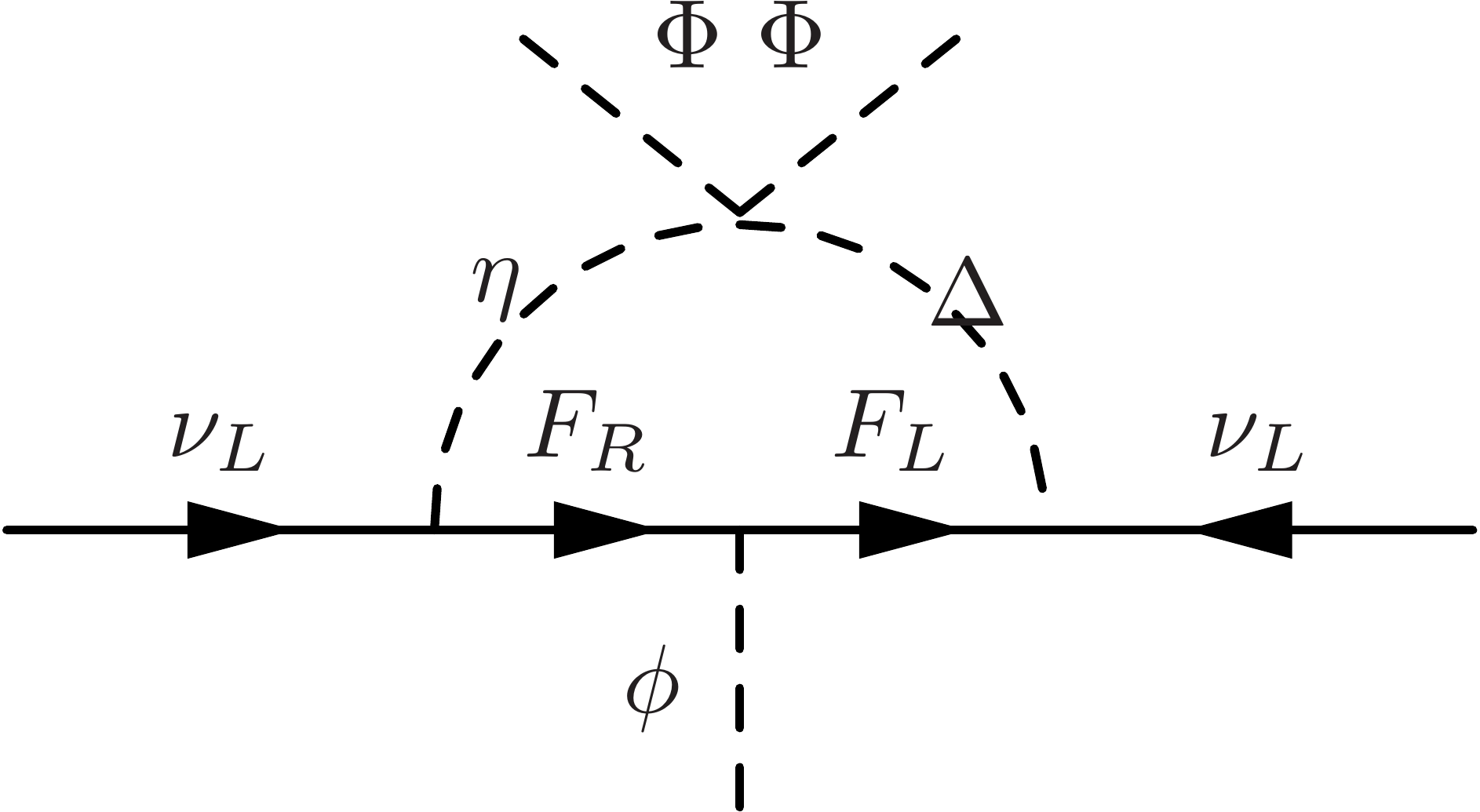}
\caption{Feynman graph for one-loop radiative neutrino mass.}
\label{Fig:neutrinomass}
\end{figure}

The scalar doublet $\Phi$ contains three would-be Goldstone bosons that become the longitudinal components of the weak gauge bosons. Its remaining degree of freedom mixes with the singlet $\phi$ into a physical neutral scalar $H$ of mass $m_H=v\sqrt{2\lambda_\Phi-\lambda_{\Phi\phi}}$, which is identified with the discovered $125~\GeV$ scalar, and a physical scalar $h$, which will only gain a radiative mass. Employing the result from Ref.~\cite{Gildener:1976ih}, we obtain
\begin{equation}
m_{h}^2=8B\langle\varphi\rangle^2,
\end{equation}
where $\langle\varphi\rangle^2=v^2+u^2=-m_H^2/\lambda_{\Phi\phi}$, and
\begin{eqnarray}
B&=& \frac{1}{64\pi^2\langle\varphi\rangle^4}
\big[\textrm{Tr}M_S^4+3\textrm{Tr}M_V^4
-4\textrm{Tr}M_F^4\big],
\end{eqnarray}
which sums over the masses of all scalars, gauge bosons and fermions. The new scalars must be sufficiently heavy to make $B>0$. The scalar singlet $\eta$ mixes with the neutral component $\Delta^0$ of the triplet scalar through the $\lambda_{\eta\Delta\Phi}$ coupling. Since the coupling also enters the radiative neutrino mass as shown in Fig.~\ref{Fig:neutrinomass}, it should be naturally small. The mass splitting among the components $\Delta^{++,+,0}$ is controlled by the $\lambda_{\Phi\Delta 2}$ coupling. As we do not discuss cascade decays between those components in this work, we simply assume it vanishes. Thus,
\begin{equation}
m_\Delta^2=\frac{1}{2}\left(\lambda_{\Phi\Delta 1} v^2 +\lambda_{\phi\Delta} u^2\right),\qquad
m_\eta^2=\frac{1}{2}\left(\lambda_{\Phi\eta} v^2+\lambda_{\eta\phi} u^2\right).
\end{equation}

\begin{figure}[!h]
	\centering
	\includegraphics[width=0.6\textwidth]{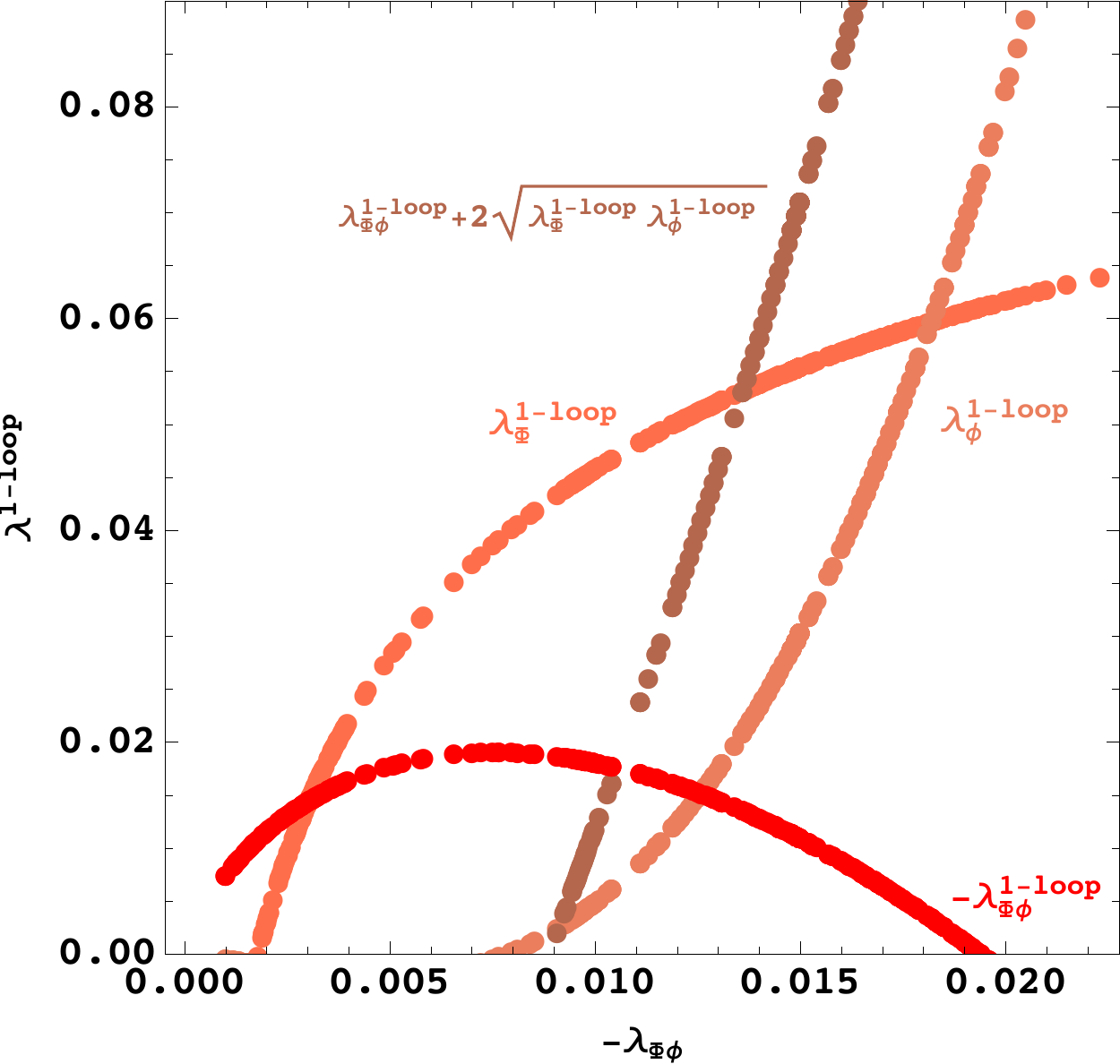}
\caption{One-loop corrected couplings are shown as functions of $|\lambda_{\Phi\phi}|$ at $m_{\eta}=200~\GeV$.}
\label{Fig:vc_stblt}
\end{figure}

The vacuum stability demands the couplings to satisfy the conditions
\begin{equation}
\lambda_{\Phi}^{\rm{1-loop}} > 0, \ \lambda_{\phi}^{\rm{1-loop}} > 0, \ \  \lambda_{\Phi\phi}^{\rm{1-loop}} + 2\sqrt{\lambda_{\Phi}^{\rm{1-loop}}\lambda_{\phi}^{\rm{1-loop}}} > 0,
\end{equation}
and the existence of two nonvanishing VEVs requires  $\lambda_{\Phi\phi}^{\rm{1-loop}} < 0$. Here the $\lambda^{\rm{1-loop}}$ couplings have included the one-loop corrections, and are defined by partial derivatives of the effective potential in the same form as appearing in, e.g., Ref.~\cite{Ahriche:2016cio}. In Fig.~\ref{Fig:vc_stblt}, they are shown as functions of $|\lambda_{\Phi\phi}|$ at $m_\eta = 200~\GeV$. It can be seen that $\lambda_{\Phi\phi}$ must lie in the range $(-0.02,-0.009)$ to satisfy the conditions.

Finally, the neutrino mass matrix in Fig.~\ref{Fig:neutrinomass} is calculated as~\cite{Law:2013saa}
\begin{equation}
{\cal M}_\nu^{\alpha\beta}=\frac{\sqrt{2}}{\left(4\pi\right)^2}
\frac{\lambda_{\eta\Delta \Phi}v^2}{m_\Delta+m_\eta} \sum_i\Big(y_\Delta^{\alpha i}y_\eta^{i\beta}
+y_\Delta^{\beta i}y_\eta^{i\alpha}\Big)
\frac{R^i_\eta R^i_\Delta}{R^i_\eta-R^i_\Delta}
\bigg(\frac{\ln R^i_\eta}{1-(R^i_\eta)^2}
-\frac{\ln R^i_\Delta}{1-(R^i_\Delta)^2}\bigg),
\end{equation}
where $R^i_{\Delta,\eta}=m_{N_i}/m_{\Delta,\eta}$. In the basis where the charged leptons are already diagonalized,  diagonalization $U^T_\textrm{PMNS}{\cal M}_\nu U_\textrm{PMNS}=m_\nu$ will yield the neutrino masses $m_{\nu_i}$ and the PMNS matrix $U_\textrm{PMNS}$. Although the desired order of magnitude for neutrino masses can be always realized by adjusting jointly the new couplings and masses, we are interested in the region of parameter space where some new particles would in principle be reachable at the LHC, i.e., with a mass not exceeding few TeV; e.g., assuming $(m_\Delta,m_\eta,m_N)\sim (10^3,200,650)~\GeV$ and $y_\Delta\sim y_\eta\sim 10^{-3},~\lambda_{\eta\Delta \Phi}\sim 10^{-4}$, will yield $\mathcal{O}\big(0.1~\eV\big)$. With generally complex $3\times n_F$ matrix $y_\Delta$ and $n_F\times 3$ matrix $y_\eta$ and a diagonal $n_F\times n_F$ real matrix formed with $R^i_{\Delta,\eta}$, where $n_F$ is the number of new doublet fermions, it is also easy to accommodate the measured PMNS matrix with free Dirac and Majorana CP violation phases. In our phenomenological analysis, we assume $n_F=3$ almost degenerate doublet fermions, although it would be enough to generate two non-vanishing neutrino masses with two fermions.

The lepton-flavor-violating (LFV) processes generically take place in neutrino mass models. Currently, the most stringent experimental bounds are set in the $\mu-e$ sector, with ${\rm{Br}}(\mu\to e\gamma)<4.2\times 10^{-13}$~\cite{TheMEG:2016wtm} and ${\rm{Br}}(\mu\to eee)<1.0\times 10^{-12}$~\cite{PDG18} for the decays and ${\rm{Br}}(\mu{\rm{Au}}\to e{\rm{Au}})<7.0\times 10^{-13}$~\cite{Bertl:2006up} for $\mu-e$ conversion in nuclei. For the model under consideration, these processes appear at the one-loop level. We have calculated the branching ratios, and found that those bounds can be readily avoided, e.g., at the benchmark point of masses of this study, namely $m_\Delta\sim 1~\TeV,~m_\eta\sim 200~\GeV,~m_F\sim650~\GeV$, assuming no special flavor structure for the Yukawa matrices $y_\Delta\sim y_\eta\sim y$, we found that a loose bound $y\lesssim\mathcal{O}(0.1)$ can satisfy all of the above LFV constraints.

\section{Dark Matter Direct Detection}\label{sec:DM}

The lightest $Z_2$-odd neutral particle could in principle serve as a DM particle. However, to avoid strong constraints from direct detection, it should preferably not couple to the $Z$ boson. As a matter of fact, such a DM candidate would otherwise be heavier than 2.5~TeV~\cite{Law:2013saa,Araki:2011hm,Araki:2010zz}, making detection of all $Z_2$-odd particles essentially impossible at the LHC. We thus choose to work with the scalar singlet $\eta$ as DM and expect to have viable parameter space for a DM mass at the electroweak scale.

The DM annihilation proceeds predominantly through the $s$-channel $H/h$ exchange and contact interactions involving $H/h$ as in a Higgs-portal DM model. The contribution from the $t-$channel exchange of the fermion $F$, which results in the $\ell\ell^\prime,~\nu\nu^\prime$ final states, is negligible as it is suppressed by the small Yukawa couplings entering the neutrino mass. The set of relevant couplings is thus $(\lambda_\Phi,~\lambda_\phi,~\lambda_{\Phi\phi},
~\lambda_{\Phi\eta},~\lambda_{\eta\phi})$. The fixed mass of $m_H=125~\GeV$ and the flat direction condition provide two constraints on $\lambda_\Phi,~\lambda_\phi$ and $\lambda_{\Phi\phi}$. The mixing of $\phi$ and $\Phi$ into $H$ and $h$ is determined by the angle $\theta$ with $\sin^2\theta=\lambda_{\Phi\phi}/
(\lambda_{\Phi\phi}-2\lambda_\Phi)=-\lambda_{\Phi\phi}v^2/m_H^2$. The angle $\theta$ enters the $\eta$ pair interactions with the $H$ and $h$ fields through $\lambda_{\Phi\eta}$ and $\lambda_{\eta\phi}$ terms. As we will see shortly, the interference effects between $H$ and $h$ are already rich enough with the $\theta$ angle and one of the two couplings $\lambda_{\Phi\eta},~\lambda_{\eta\phi}$. We therefore technically switch off one of them, say $\lambda_{\Phi\eta}=0$, to reduce the number of parameters. This leaves us with two free parameters, which we choose to be $(\lambda_{\Phi\phi},~\lambda_{\eta\phi})$, or equivalently $(\lambda_{\Phi\phi},~m_{\eta})$.

\begin{figure}
\centering
	\includegraphics[scale=.85]{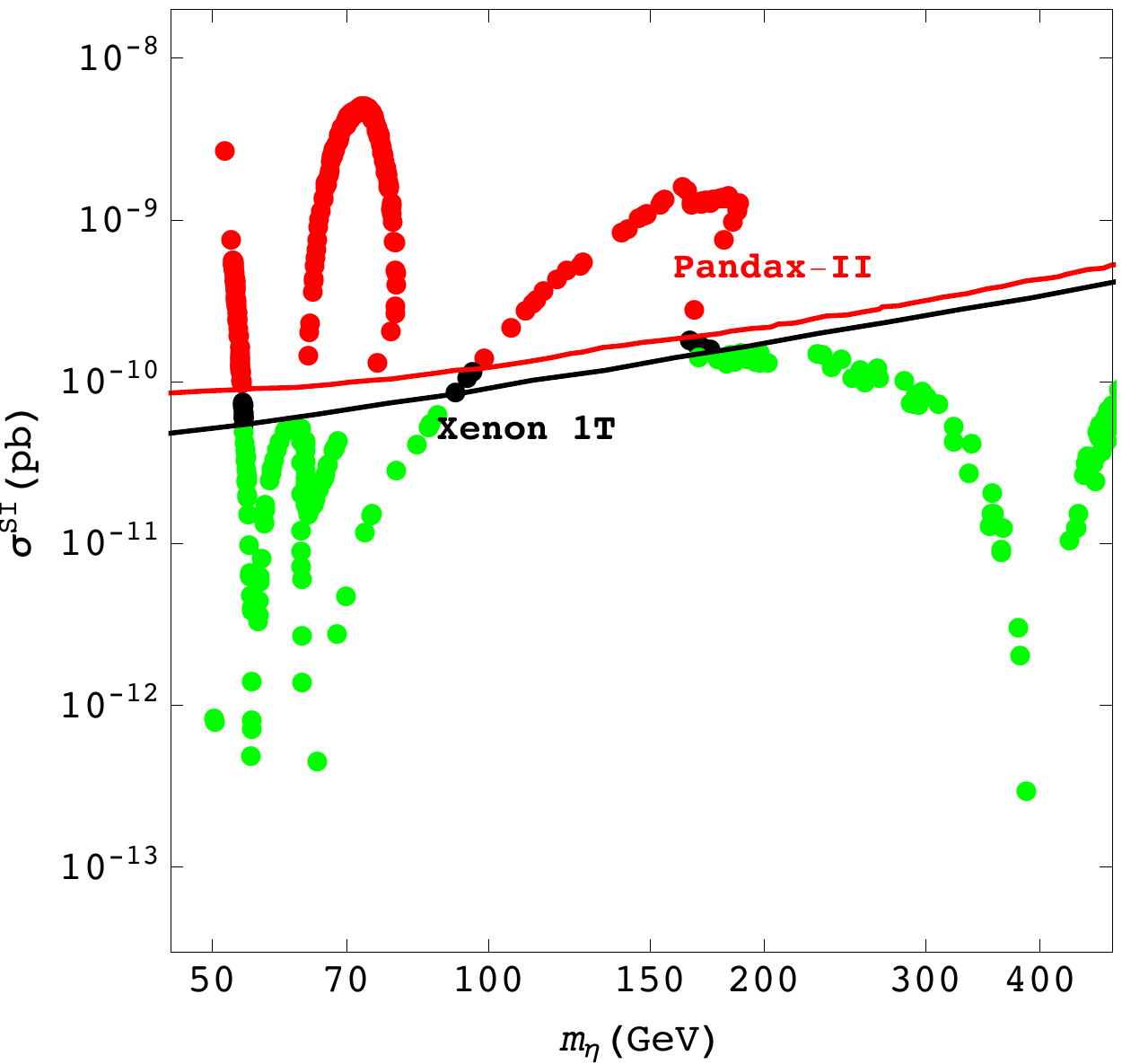}
\caption{Sampled spin-independent cross section for DM scattering off the nucleon as a function of DM mass. The two curves are the upper bounds by the PandaX-II~\cite{Cui:2017nnn} and Xenon1T~\cite{Aprile:2018dbl} experiments.}
\label{fig:DD}
\end{figure}

\begin{figure}
\centering
\includegraphics[scale=1]{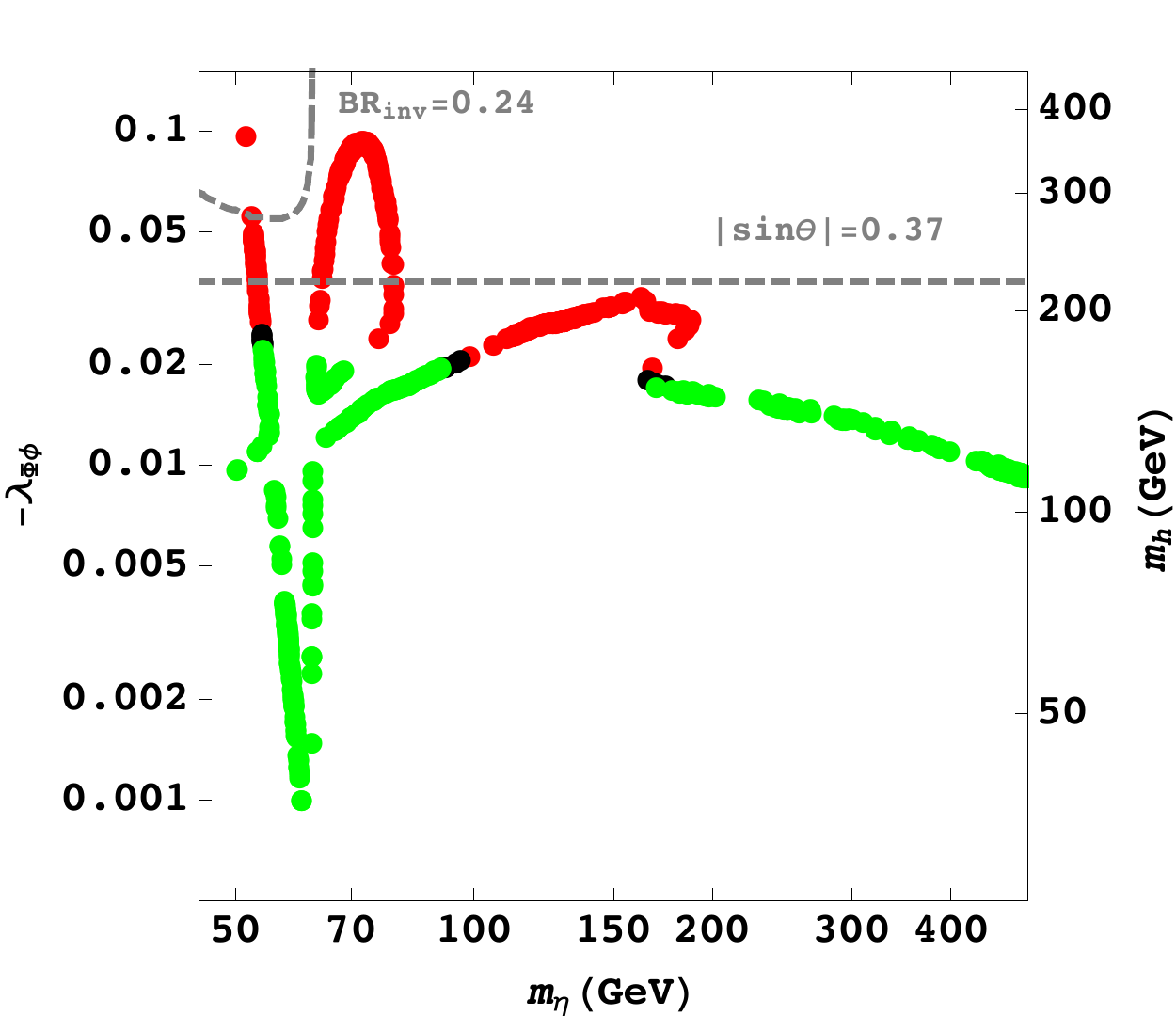}
\caption{Scanning results from DM simulation projected on the coupling $-\lambda_{\Phi\phi}$ and $m_h$ against DM mass $m_\eta$. The horizontal dashed line indicates the upper bound on $|\sin\theta|$, and the dashed curve shows the branching ratio upper bound on invisible Higgs decay.}
\label{fig:DMrd}
\end{figure}

In our numerical analysis, we use the package \texttt{micrOMEGAs}~\cite{Belanger:2018mqt} to calculate the DM relic density and cross section for DM-nucleon scattering. We scan the two free parameters in the following ranges
\begin{equation}
|\lambda_{\Phi\phi}|\in(0.0001,0.1),\ \ m_{\eta}\in(45,500)~\GeV,
\end{equation}
while the other new particle masses are fixed as
\begin{equation}
m_\Delta=1000~\GeV, \ \ m_F=650~\GeV.
\end{equation}
Our numerical results are shown as a function of the DM mass $m_\eta$ in Fig.~\ref{fig:DD} for the spin-independent cross section $\sigma^\textrm{SI}$ in DM direct detection and in Fig.~\ref{fig:DMrd} for the coupling $-\lambda_{\Phi\phi}$ (left longitudinal axis) and $m_h$ (right longitudinal axis). In these figures, all displayed points pass the DM relic density requirement. The red points are excluded by the PandaX-II experiment~\cite{Cui:2017nnn}, the black ones are further excluded by the Xenon1T result~\cite{Aprile:2018dbl}, and the green ones are still allowed by all current experimental data.

Two competing physical effects are responsible for the shapes of the points in Figs.~\ref{fig:DD} and \ref{fig:DMrd} and in particular for the jitter behaviour in the low DM mass region. The first is the destructive interference between the scalars $H$ and $h$ in the DM annihilation (via $s$-channel exchanges) and direct detection ($t$-channel) amplitudes. They contribute a similar term but differ in sign, because the relevant couplings are proportional to $\pm\sin\theta\cos\theta$. This effect is strong when the mass $m_h$ of $h$ approaches $m_H$ of the SM-like Higgs $H$. The second effect is the on-resonance enhancement of DM annihilation when the DM mass $m_\eta$ approaches half of the scalar mass $m_H$ or $m_h$. The first effect takes place without requiring the couplings themselves to be necessarily small, and it is evident due to the presence of a dip in $\sigma^\textrm{SI}$ at $m_\eta\lesssim 400~\GeV$, where $m_h$ crosses $m_H$, but in the absence of a corresponding dip in $-\lambda_{\Phi\phi}$. The second effect would appear as a sudden and deep dip at $m_\eta\approx m_H/2$, as in the usual case, when there is a single Higgs portal $H$. This then requires the relevant couplings to be very small to avoid over-annihilation of DM. However, the presence of a second scalar $h$ makes the situation highly involved, especially in the region $m_\eta\sim 50-70~\GeV$, where $m_h$ varies rapidly and moves across the value of $m_H$. In this region, the destructive interference and successive appearance and disappearance of one or two resonances overlap to varying extents, and also explain the unusual rising behaviour as $m_\eta$ increases from $\sim 70~\GeV$ to $\sim 100~\GeV$.

As mentioned above, the $s$-channel annihilation through the exchange of $H/h$ is part of the dominant DM annihilation channels. In the low mass region DM annihilates mainly into $b\bar{b}$ and $W^+W^-$ final states through $s$-channel exchange, while in the high mass region it annihilates dominantly into the $HH,~hh,~Hh$ particles. The transition point appears around $m_\eta\in(160,200)~\GeV$, where one sees a drop of $\lambda_{\Phi\phi}$ as the contact annihilation channels are opened and start to contribute. Comparing to Ref.~\cite{Ahriche:2016cio}, which excluded the mass window of a singlet fermion DM from $10$~GeV to $200$~GeV, here we find that a significant portion of the mass range is still viable for a singlet scalar DM.

When DM is light enough, $m_\eta<m_{H}/2$, it contributes to the invisible decay of the Higgs boson. The decay width is given by \begin{equation}
\Gamma(H\to \eta\eta)=
\frac{\lambda_{\Phi\phi}^2 v^2}{8\pi\left(m_H^2 + \lambda_{\Phi\phi} v^2\right)} \frac{m_\eta^4}{m_{H}^3}
\sqrt{1-\frac{4m_\eta^2}{m_{H}^2}}.
\end{equation}
The latest searches for invisible Higgs decays by the CMS based on the $5.1,~19.7,~2.1~\invfb$ data collected at $7,~8,~13~\TeV$, respectively, set a combined bound on the invisible branching ratio $\rm{Br}_{\rm{inv}}<0.24$~\cite{Khachatryan:2016whc} in the production modes of $gg$F, VBF, $ZH$, and $WH$. At the ATLAS, the most stringent bound comes from the study on the $8~\TeV$~$20.3~\invfb$ data through the VBF production, $\rm{Br}_{\rm{inv}}<0.28$~\cite{Aad:2015txa}. It is easy to check that invisible decays are currently less restrictive than direct detection. Finally, the mixing between the $\Phi$ and $\phi$ fields suppresses all the couplings between the Higgs boson $H$ and SM particles. Using the data on direct searches of the Higgs boson, an upper bound $|\sin\theta |<0.37$ has been achieved at $95\%$~C.L.~\cite{Farzinnia:2013pga}. We can see from Fig.~\ref{fig:DMrd} that our survived sample points safely avoid this constraint.

\section{LHC PHENOMENOLOGY}\label{sec:LHC}

As discussed in section~\ref{sec:MD}, the scalar particles must be sufficiently heavy to guarantee the expected spontaneous symmetry breakdown. In contrast, we have chosen the singlet scalar as DM to avoid strong constraints from direct detection. Thus, a natural order of masses for the $Z_2$ odd particles suggests itself:
\begin{equation}
m_{\Delta} > m_F > m_{\eta}
\label{eq_massorder}
\end{equation}
The $Z_2$ symmetry implies that a heavier $Z_2$-odd particle decays into a lighter one plus SM particles. Since we have assumed a degenerate spectrum for the triplet scalars, the permitted decays are $\Delta^{++}\to E^+\ell^+,~\Delta^+\to (E^+\nu,~N\ell^+),~E^+\to\eta\ell^+$ plus their conjugates, where $\ell$ is a charged lepton. These decays generate events with multiple charged leptons at LHC, while the pure neutral decays $\Delta^0\to N\nu$ and $N\to\nu\eta$ will appear as missing energy.

Now, we focus on searching for the LHC signatures of the new particles and interactions. We follow the standard procedure for event simulation and analysis using a series of software programs. We utilize {\tt FeynRules}~\cite{Christensen:2008py} to obtain the {\tt UFO}~\cite{Degrande:2011ua} model file, which is input into {\tt MadGraph\_aMC@NLO}~\cite{Alwall:2011uj} to generate parton level events. The {\tt NNPDF2.3}~\cite{Ball:2012cx} LO parton distribution function set passing through {\tt Pythia 6}~\cite{Sjostrand:2006za} is used to generate showering and hadronization, successively. The events then pass {\tt Dephes3}~\cite{Ovyn:2009tx} for detector simulation. After simulation, we use {\tt MadAnalysis5}~\cite{Conte:2012fm} to obtain various final-state distributions for analysis. Finally, {\tt CheckMATE}~\cite{Drees:2013wra} is used to examine whether the benchmark points we choose are excluded or not at 95\% C.L.

\begin{figure}[!htbp]
	\centering
	\includegraphics[width=0.49\linewidth]{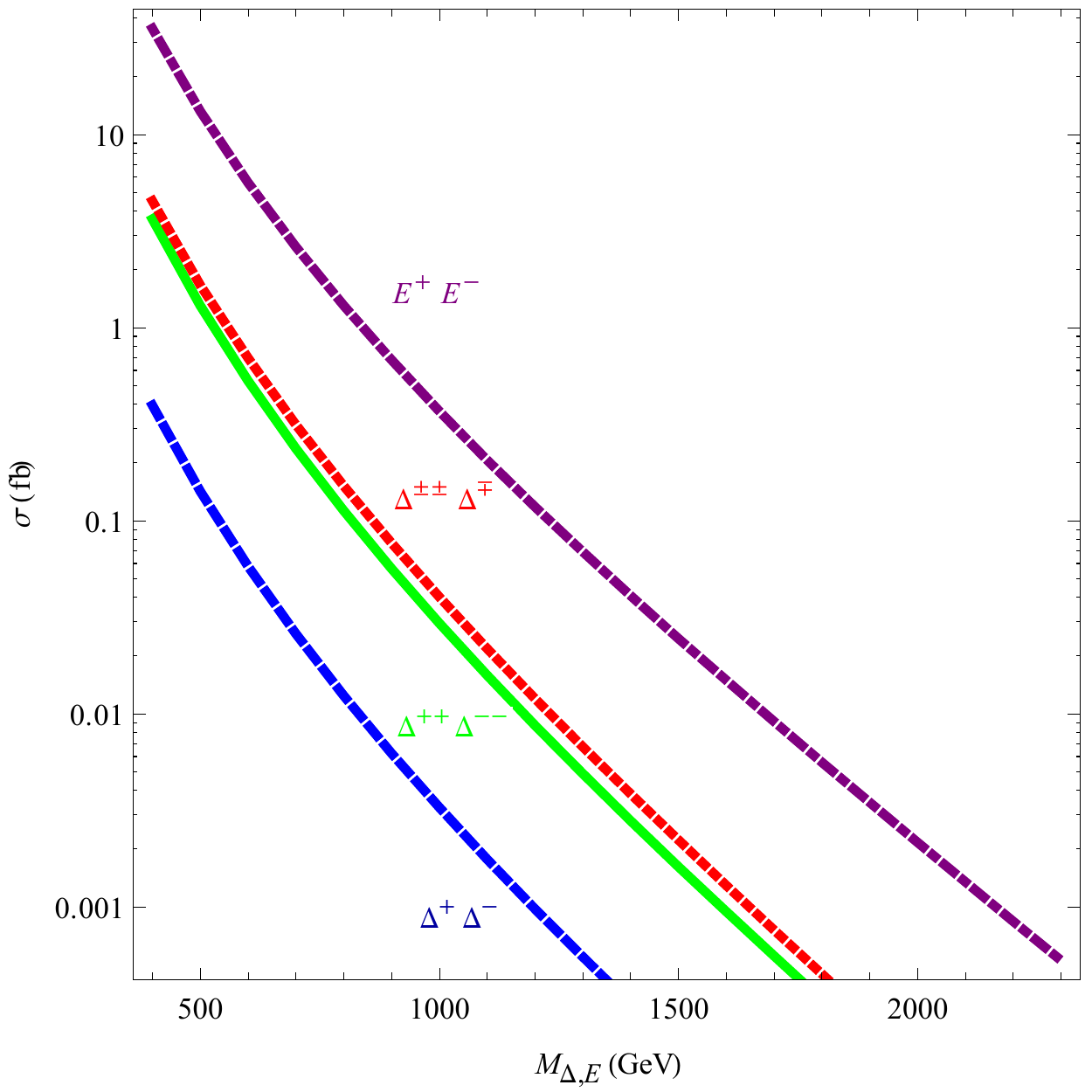}
	\caption{Cross sections for pair and associated production of $\Delta^{++}\Delta^{--}$, $\Delta^{\pm\pm}\Delta^{\mp}$, $\Delta^+\Delta^{-}$ and $E^+ E^-$ at 13\ TeV LHC as a function of their masses $m_\Delta$ and $m_{E^{\pm}}$.}
	\label{fig:xsec13}
\end{figure}

Let us first take a look at the production of heavy particles through electroweak interactions. The cross sections for the dominant pair and associated production of the scalar triplet and fermion doublet are shown in Fig.~\ref{fig:xsec13} at 13~TeV LHC. The production of $E^{+}E^{-}$, $\Delta^{\pm\pm}\Delta^{\mp}$, and $\Delta^{++}\Delta^{--}$ will lead respectively to signatures of dilepton, trilepton, and four-lepton with a large missing transverse energy $\slashed E_T$ carried away by the DM particle $\eta$. We discard the dilepton events produced via $\Delta^{+}\Delta^{-}$ production, because its cross section is too small compared with $E^{+}E^{-}$. At the same time, we combine the trilepton and four-lepton signatures (termed multi-lepton below) to enhance the significance. To summarize, the following two signatures will be studied in detail:
\begin{itemize}
	\item $2\ell^\pm + \slashed{E}_T$ from $E^{+}E^{-}$ production,
	\item $3\ell^\pm + \slashed{E}_T$ and $4\ell^\pm + \slashed{E}_T$ from $\Delta^{\pm\pm}\Delta^{\mp}$ and $\Delta^{++}\Delta^{--}$ production,
\end{itemize}
where $\ell = e,~\mu$ in our definition of a lepton for LHC signatures.

The search for dilepton and multi-lepton plus $\slashed{E}_T$ signatures has recently been performed at 13~TeV LHC by ATLAS~\cite{Aaboud:2018jiw} based on simplified SUSY models. Owing to model dependence, its exclusion line has to be recast for the model under consideration. Fixing the DM mass at 200~GeV and according to entire cuts in~\cite{Aaboud:2018jiw}, we have used {\tt CheckMATE} to perform this recasting job, and have confirmed that heavy leptons are excluded at a mass of 600~GeV. Thus, we choose $m_E = 650~\GeV$. Considering the order of masses in eq. (\ref{eq_massorder}) and the experimental lower bound on $m_{\Delta}$~\cite{Aaboud:2017qph}, we work with the following benchmark point to illustrate the testability of signatures at LHC:
\begin{equation}
m_\Delta = 1000\ {\rm GeV}, m_E = 650\ {\rm GeV}, m_\eta = 200\ {\rm GeV}.
\label{eqn_bmp1}
\end{equation}
Since the current 36.1/fb data at 13~TeV LHC has a small significance for the signals under consideration, we present our results for the future HL-LHC with an integrated luminosity of 3000/fb at 13~TeV.

\subsection{Dilepton Signature}

The dilepton signature from the pair production of $E^{\pm}$ is as follows:
\begin{equation}
pp \to E^+ E^- \to \eta \ell^+ \eta \ell^- \to \ell^+ \ell^- + \slashed{E}_T,
\end{equation}
where $\ell = e,~\mu$ for collider simulations. Requiring clean backgrounds, we concentrate on the final states as ATLAS did for the dilepton signature. The dominant sources of background are di-bosons ($WZ,~ZZ,~WW$), tri-bosons ($VVV$ with $V = W,~Z$), top pairs ($t\bar t$), and top plus boson (mainly from $t\bar tV$) with leptonic decays of $W,~Z$, all of which we have included at the leading order, i.e., without a $K$ factor. We adopt the same selection criteria as ATLAS for a more reasonable analysis.

\begin{figure}[!htbp]
	\begin{center}
		\includegraphics[width=0.49\linewidth]{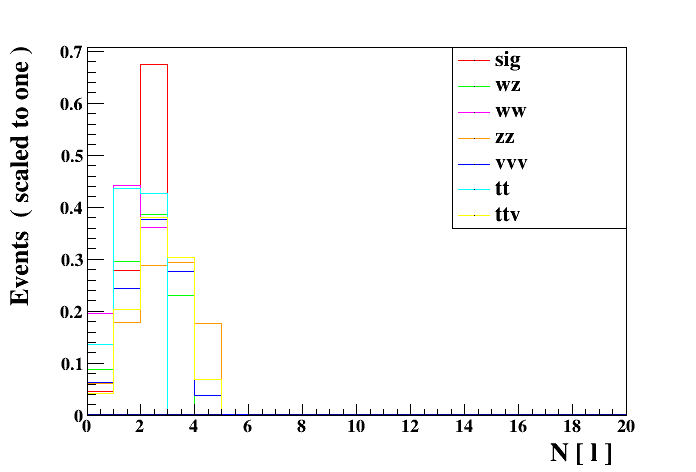}
		\includegraphics[width=0.49\linewidth]{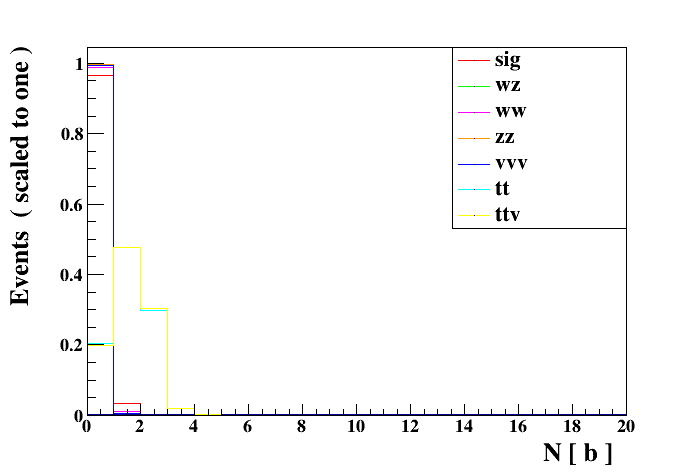}
	\end{center}
	\caption{Distributions of $N(\ell)$ and $N(b)$ at 13\ TeV LHC for dilepton signature.}
	\label{fig:bscts_dilep}
\end{figure}

\begin{figure}[!htbp]
\begin{center}
\includegraphics[width=0.49\linewidth]{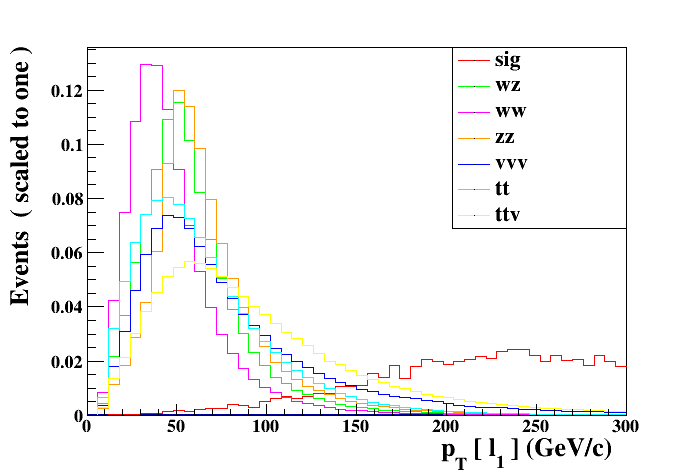}
		\includegraphics[width=0.49\linewidth]{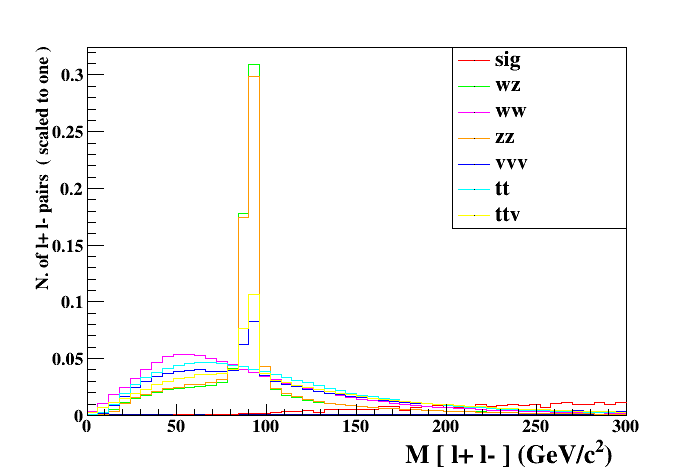}
		
\includegraphics[width=0.49\linewidth]{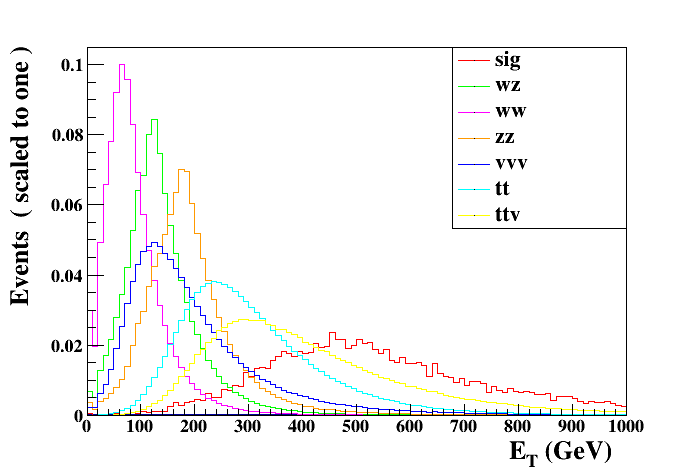}
		\includegraphics[width=0.49\linewidth]{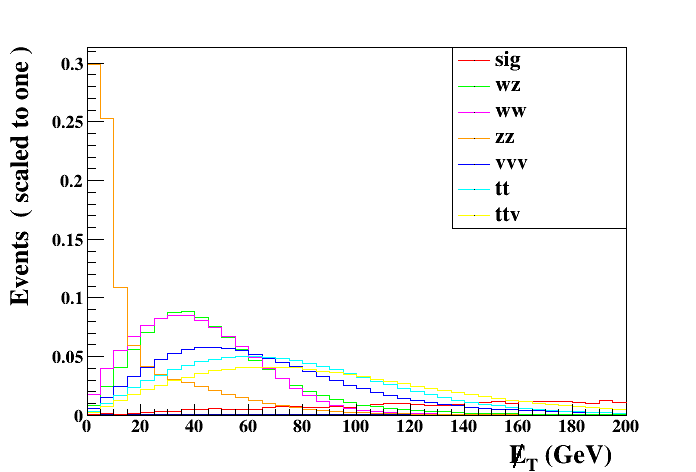}
\end{center}
\caption{Distributions of $p^{\ell_1}_T$, $m_{\ell\ell}$, $E_T$, and\ $\slashed E_T$ at 13~TeV LHC for dilepton signature.}
	\label{fig:cuts_dilep}
\end{figure}

We start with some basic cuts:
\begin{equation}
p^{\ell_1}_T > 25\ {\rm GeV},~\ p^{\ell_2}_T > 20\ {\rm GeV}, ~m_{\ell \ell} > 40\ {\rm GeV},\ |\eta| < 2.47,
\label{eq: basic cuts}
\end{equation}
where $p^{\ell_1(\ell_2)}_T$ denotes the transverse momentum of the more (less) energetic one in the two charged leptons, $m_{\ell \ell}$ is the dilepton invariant mass, and $\eta$ is the pseudorapidity. In Fig.~\ref{fig:bscts_dilep}, the distributions of the numbers of leptons $N(\ell)$ and bottom-quark jets $N(b)$ at 13~TeV are shown. In order to make backgrounds cleaner, we apply the following cuts:
\begin{equation}
N({\ell}^+) = 1~{\rm and}\ N({\ell}^-) = 1;~N(b) = 0,
\end{equation}
which require exactly a pair of oppositely charged leptons and no appearance of a $b$-jet to cut the backgrounds coming from $t \bar t$ and $t \bar t V$ production. Using {\tt Madanalysis5}, we obtain other important distributions shown in Fig.~\ref{fig:cuts_dilep}. As the large DM and fermion masses make the signal deviate significantly from the backgrounds in the $p^{\ell_1}_T$ and $\slashed E_T$ distributions, we impose further cuts on them:
\begin{equation}
p^{\ell_1}_T > 250~\GeV,~\slashed E_T > 200\GeV.
\end{equation}

In table~\ref{table: di-cutflow} we show the cut-flow for the dilepton signature at our main benchmark point in eq.~(\ref{eqn_bmp1}) and for dominant backgrounds. Also included are two more benchmark points with slightly larger masses ($m_\Delta=1200~\GeV$ and $m_E=700,~750~\GeV$), for the purpose of comparison. Although the triplet scalar does not directly affect the dilepton signal a larger mass helps stabilize the vacuum when the fermion mass increases. The efficiency of the $p^{\ell_1}_T$ and $\slashed E_T$ cuts is clearly appreciated, while the $t\bar t$ and $t\bar tV$ backgrounds are significantly reduced by demanding $N(b)=0$. As expected, a larger fermion mass $m_E$ results in smaller cross section and signal events. After all the cuts, we have about $1601.1$, $1367.7$, and $1020.3$ signal events at the three benchmark points, respectively.

\begin{table}
\begin{center}
\begin{tabular}{|c|c|c|c|c|c|c|c|c|}
			\hline
			Channels   &     No cuts & Basic cuts &
			$N(\ell) = 2$   &$N(b) = 0$&$p^{\ell_1}_T > 250\ {\rm GeV}$& $\slashed E_T > 200\ {\rm GeV}$\\
			\hline
			
			1000-650-200 &   4653&  3132.9& 3103.1& 3015.6 & 2531.7 &   1601.1 \\
			1200-700-200 &  3294 & 2073.9 & 2055.8 & 1994.2 & 1688.2 & 1367.7 \\
			1200-750-200 & 2300 & 1446.3 &1432.0 & 1386.7 &1211.0 & 1020.3 \\
			
			\hline
			
			$WZ$    &   1085099&  472116& 228205& 227192 & 1395.4 & 229\\
			\hline
			
			$WW$     &   8405999&  2083716  & 2075393 & 2059144 & 19472 & 444.1\\
			\hline
			
			$ZZ$     &   126959&  79381  & 20136& 20047 & 198.3 & 6.35\\
			\hline
			
			$VVV$     &   3963&  2015.9  & 817.4& 812.2 & 46.57 & 5.22\\
			\hline
			
			$t \bar t$     &   68129999&  21391431  & 20167294& 3943440 & 70014 & 1936.6\\
			\hline
			$t {\bar t} V$     &   12984&  8170.6 & 2313.7& 430.9 &  29.80 & 2.98\\
			\hline
		\end{tabular}
\end{center}
\caption{Cut-flow for dilepton signature at benchmark points and dominant backgrounds at 13~TeV LHC with integrated luminosity of 3000~${\rm fb}^{-1}$. Each set of numbers in the first column shows the values of $m_{\Delta,E,\eta}$ in GeV at each benchmark point.}
\label{table: di-cutflow}
\end{table}

\subsection{Multi-lepton Signature}

\begin{figure}[!htbp]
	\begin{center}
		\includegraphics[width=0.45\linewidth]{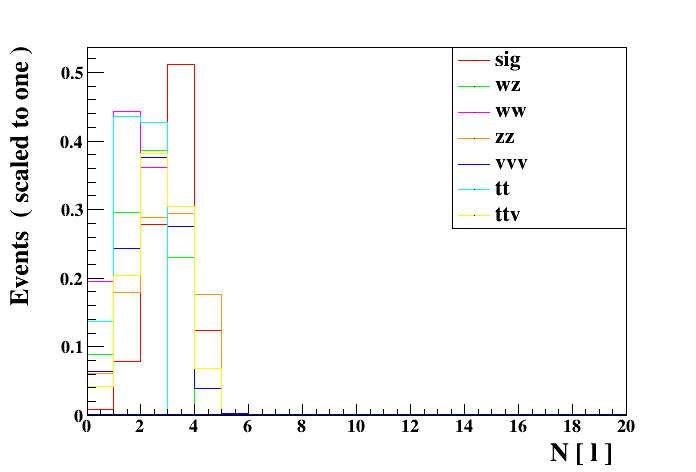}
		\includegraphics[width=0.45\linewidth]{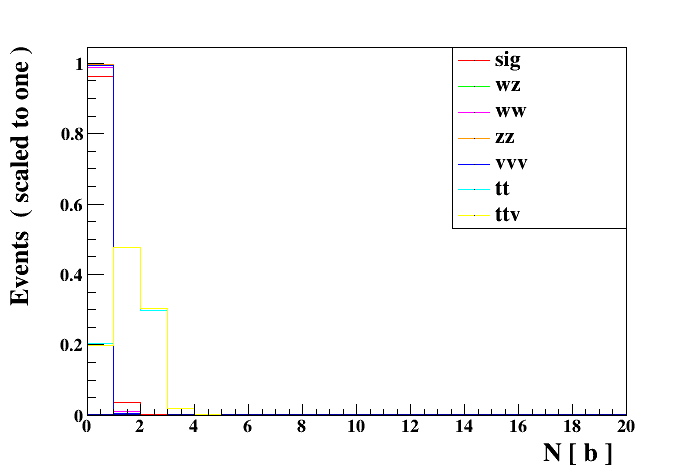}
		\includegraphics[width=0.45\linewidth]{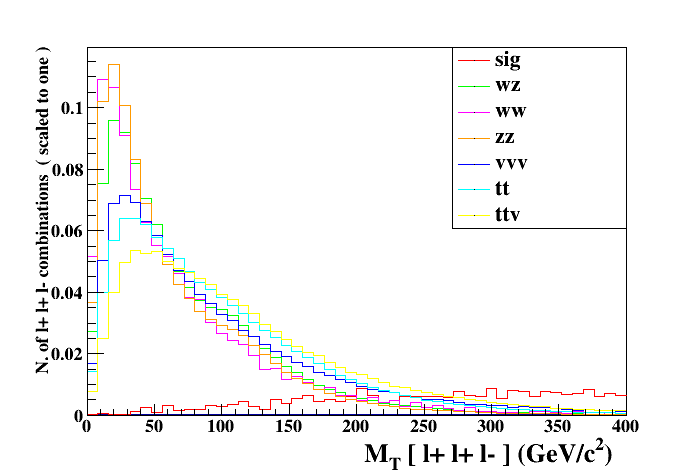}
		\includegraphics[width=0.45\linewidth]{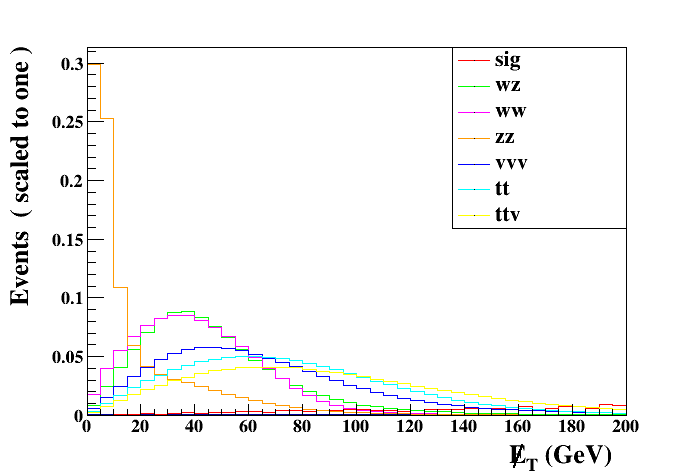}
		\includegraphics[width=0.45\linewidth]{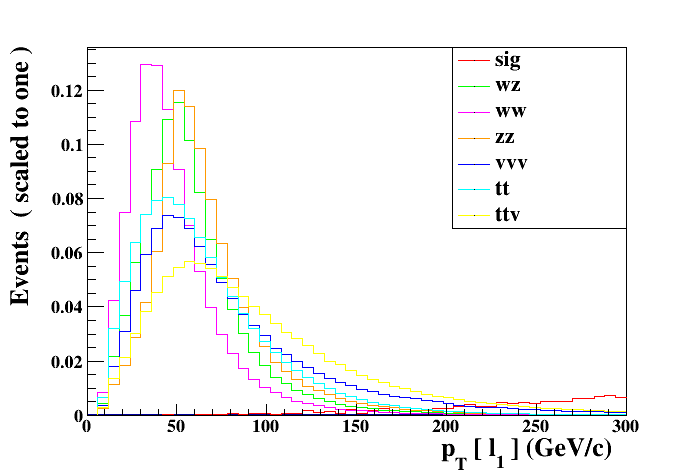}
		\includegraphics[width=0.45\linewidth]{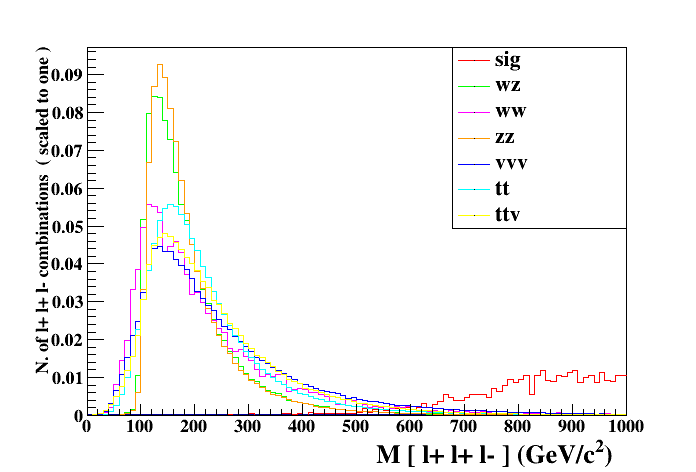}
		
	\end{center}
\caption{Distributions at 13~TeV LHC for multi-lepton signature of $N(\ell)$, $N(b)$, $\slashed E_T$, $M_T(\ell^+\ell^+\ell^-)$, $M(\ell^+\ell^+\ell^-)$, and $p^{\ell_1}_T$.}
	\label{fig: multi-cuts}
\end{figure}

The multi-lepton signature originates from the production of $\Delta^{++}\Delta^{--}$ and $\Delta^{\pm\pm}\Delta^{\mp}$ and their sequential decays:
\begin{equation}\begin{split}
pp &\to \Delta^{++}\Delta^{--}
\to E^+ \ell^+ E^- \ell^-
\to\eta\ell^+\ell^+\eta\ell^-\ell^-
\to 2\ell^+ 2\ell^- + \slashed E_T,
\\
pp &\to \Delta^{\pm \pm} \Delta^{\mp}
\to E^{\pm} \ell^{\pm}~E^{\mp}\nu/N \ell^{\mp}
\to\eta\ell^{\pm}\ell^{\pm}~\eta\ell^{\mp}\nu/\eta\nu \ell^{\mp}
\to 2\ell^{\pm}\ell^{\mp} + \slashed E_T.
\end{split}\end{equation}
We start again with the basic cuts in Eq.~(\ref{eq: basic cuts}), then we select the trilepton and four-lepton events by imposing the following criteria:
\begin{equation}
N(\ell^{\pm})=2\textrm{ and }N(\ell^{\mp})=1,
\textrm{ or }N(\ell^+)=2\textrm{ and }N(\ell^-)=2;~
N(b) = 0.
\end{equation}
We demand that the trilepton events contain no like-sign trileptons, and that in the four-lepton events there are exactly two positively and two negatively charged leptons. From the possibly relevant distributions shown in Fig.~\ref{fig: multi-cuts}, we propose the stricter cuts:
\begin{equation}
p^{\ell_1}_T>350~\GeV,~\slashed E_T>250~\GeV,~M(\ell^+ \ell^+\ell^-)>800~\GeV.
\end{equation}

Table~\ref{table: multi-cutflow} shows the cut-flow for the multi-lepton signature at the benchmark point eq.~(\ref{eqn_bmp1}), and another two with slightly heavier triplet scalars and for the main backgrounds. The production cross section and the number of signal events decrease as $m_\Delta$ increases. Because of the relatively small cross sections for the tri- and four-lepton signal events, we provide only the results for the HL-LHC mode. The cuts we employed here are efficient enough in preserving signals while suppressing backgrounds. The $p^{\ell_1}_T$ cut is still the golden cut for the multi-lepton case for reducing a large portion of the backgrounds. After the $\slashed E_T$ cut, only the $WZ$ events in the backgrounds are significant and are further diminished by a strong cut on the invariant mass for the trilepton system $M(\ell^+\ell^+\ell^-)$. The backgrounds in the multi-lepton case are much cleaner than in the dilepton case, however at the cost of less signal events: after all the cuts, there remain only about 16, 7.9, and 5.95 signal events, respectively.

\begin{table}
\begin{center}
\begin{tabular}{|c|c|c|c|c|c|c|c|c|}
			\hline
			Channels   &     No cuts & Basic cuts &
			$N(\ell)\ {\rm cuts}$   &$N(b) = 0$&$p^{\ell_1}_T > 250\ {\rm GeV}$& $\slashed E_T > 200\ {\rm GeV}$ & $M(\ell^+ \ell^+ \ell^-) > 800\ {\rm GeV}$\\
			\hline
			
			1000-650-200 &   56.13&  55.67& 35.66& 34.57 & 30.55 &   22.97 & 16.05 \\
			1050-650-200 &  26.30 & 22.0 & 17.01 & 16.35 & 15.10 & 11.22 & 7.90 \\
			1100-650-200 & 18.8 & 15.76 &12.19 & 11.71 &11.03 & 8.27& 5.95 \\
			
			\hline
			
			$WZ$    &   1085099&  870159& 243626& 242741 & 161.7 &61.85& 5.43\\
			\hline
			
			$WW$     &   8405999&  5168236  & 23.8 & 21.57 & 0 & 0& 0\\
			\hline
			
			$ZZ$     &   126959&  1117606  & 592066& 589929 & 557.4 &7.62& 0\\
			\hline
			
			$VVV$     &   3963&  3341.3  & 1187.2& 1181.1 & 17.27 & 4.97& 1.11\\
			\hline
			
			$t \bar t$     &   68129999&  44683259  & 11039& 4633 & 7.3 & 0& 0\\
			\hline
			$t {\bar t} V$     &   12984&  7177.2 &4675.9& 934.4 &  9.6 & 2.49& 0.298\\
			\hline
		\end{tabular}
\end{center}
\caption{Cut-flow for dilepton signature at benchmark points and dominant backgrounds at 13~TeV LHC with integrated luminosity of 3000~${\rm fb}^{-1}$. Each set of numbers in the first column shows the values of $m_{\Delta,E,\eta}$ in GeV at each benchmark point.}
\label{table: multi-cutflow}
\end{table}

We summarize in Table~\ref{table: significance} our signal and background events together with their significance after all the cuts for both di- and multi-lepton cases. The multi-lepton signal is much cleaner than the dilepton one in both aspects of signals and backgrounds, however its significance is much less than the latter. This leaves the dilepton signal as a better search channel for the future HL-LHC.

\begin{table}
	\begin{center}
		\begin{tabular}{|c|c|c|c|c|}
			\hline
			\multicolumn{2}{|c|}{Benchmark points} & S & B & $ S/\sqrt{S+B} $\\ \cline{3-5}
			
			\hline
			\multirow{3}*{dilepton case}  & 1000-650-200 & 1601.1 & 2624.25 & 24.6 \\ \cline{2-5}
			& 1200-700-200 & 1367.7 & 2624.2 & 21.6 \\  \cline{2-5}
			& 1200-750-200 & 1020.3 & 2624.2 & 16.9 \\
			\hline
			\multirow{3}*{multi-lepton case}  & 1000-650-200 & 16.05 & 6.84 & 3.36 \\ \cline{2-5}
			& 1050-650-200 & 7.90 & 6.83 & 2.06 \\  \cline{2-5}
			& 1100-650-200 & 5.95 & 6.84 & 1.66 \\
			\hline
		\end{tabular}
	\end{center}
\caption{Summary of numbers of signal and background events after all cuts and significance for dilepton and multi-lepton cases.}
	\label{table: significance}
\end{table}

\section{Conclusion}\label{sec:conclusion}

We have studied the dark matter and LHC phenomenology in a scale-invariant scotogenic model, which addresses three issues beyond the standard model in one framework, i.e., neutrino mass, dark matter, and generation of the electroweak scale. We incorporated the constraints coming from dark matter relic density and direct detection and bounds from direct searches and invisible decays of the Higgs boson, and searched for viable parameter space for the most relevant parameters in the scalar potential $\lambda_{\Phi\phi},~m_\eta$. To test the model further at high energy colliders, we proposed to employ the dilepton and multi-lepton signatures and made a detailed simulation at $13~\TeV$ LHC with an integrated luminosity of $3000~\invfb$. We found that the dilepton channel, mainly due to its large cross section, is the most promising to probe in the future high-luminosity LHC run.

\section*{Acknowledgement}
CG and SYG are very grateful to Junjie Cao, Liangliang
Shang and Yuanfang Yue for their help in configuring the package \texttt{CheckMATE 2}. SYG thanks Ran Ding for useful discussions. This work was supported in part by the Grants No. NSFC-11575089 and No. NSFC-11025525, by The National Key Research and Development Program of China under Grant No. 2017YFA0402200, and by the CAS Center for Excellence in Particle Physics (CCEPP).


\begin{thebibliography}{00}

	\bibitem{Coleman:1973jx}
	S.~R.~Coleman and E.~J.~Weinberg,
	Phys.\ Rev.\ D {\bf 7}, 1888 (1973).
	doi:10.1103/PhysRevD.7.1888

\bibitem{one loop} L.~M.~Krauss, S.~Nasri and M.~Trodden,
Phys.\ Rev.\ D \textbf{84}, 016004 (2011) [arXiv:1101.5713 [hep-ph]];
M.~Aoki, S.~Kanemura and K.~Yagyu,
Phys.\ Lett.\ B \textbf{702}, 355 (2011) [Erratum-ibid.\ B \textbf{706}, 495
(2012)] [arXiv:1105.2075 [hep-ph]]; 
S.~S.~C.~Law and K.~L.~McDonald,
Phys.\ Lett.\ B \textbf{713}, 490 (2012) [arXiv:1204.2529 [hep-ph]];
P.~S.~Bhupal Dev and A.~Pilaftsis,
Phys.\ Rev.\ D \textbf{87}, 053007 (2013) [arXiv:1212.3808 [hep-ph]];
 S.~Y.~Guo, Z.~L.~Han and Y.~Liao,
Phys.\ Rev.\ D {\bf 94}, 115014 (2016)
[arXiv:1609.01018 [hep-ph]].
E.~Ma, I.~Picek and B.~Radovcic
Phys.\ Lett.\ B \textbf{726}, 744 (2013) [arXiv:1308.5313 [hep-ph]];
D.~Restrepo, O.~Zapata and C.~E.~Yaguna,
JHEP \textbf{1311}, 011 (2013) [arXiv:1308.3655 [hep-ph]];
V.~Brdar, I.~Picek and B.~Radovcic,
Phys.\ Lett.\ B \textbf{728}, 198 (2014) [arXiv:1310.3183 [hep-ph]];
H.~Okada and K.~Yagyu, 
Phys.\ Rev.\ D \textbf{89}, 053008 (2014) [arXiv:1311.4360 [hep-ph]];
S.~Kanemura, T.~Matsui and H.~Sugiyama,
Phys.\ Rev.\ D \textbf{90}, 013001 (2014) [arXiv:1405.1935 [hep-ph]];
  S.~Baek, H.~Okada and K.~Yagyu,
JHEP {\bf 1504}, 049 (2015)
[arXiv:1501.01530 [hep-ph]].
  H.~Okada, N.~Okada and Y.~Orikasa,
Phys.\ Rev.\ D {\bf 93}, 073006 (2016)
[arXiv:1504.01204 [hep-ph]].
  D.~Restrepo, A.~Rivera, M.~Sánchez-Peláez, O.~Zapata and W.~Tangarife,
Phys.\ Rev.\ D {\bf 92}, 013005 (2015)
[arXiv:1504.07892 [hep-ph]].
 W.~Wang and Z.~L.~Han,
Phys.\ Rev.\ D {\bf 92}, 095001 (2015)
[arXiv:1508.00706 [hep-ph]].
A.~Aranda and E.~Peinado,
Phys.\ Lett.\ B {\bf 754}, 11 (2016)
[arXiv:1508.01200 [hep-ph]];
  H.~Okada and Y.~Orikasa,
Phys.\ Rev.\ D {\bf 94}, 055002 (2016)
[arXiv:1512.06687 [hep-ph]].
  W.~B.~Lu and P.~H.~Gu,
JCAP {\bf 1605}, 040 (2016)
[arXiv:1603.05074 [hep-ph]].

\bibitem{two loops}M.~Aoki, S.~Kanemura, T.~Shindou and K.~Yagyu,
JHEP \textbf{1007}, 084 (2010) [JHEP \textbf{1011}, 049 (2010)]
[arXiv:1005.5159 [hep-ph]]; 
M.~Lindner, D.~Schmidt and T.~Schwetz,
Phys.\ Lett.\ B \textbf{705}, 324 (2011) [arXiv:1105.4626 [hep-ph]];
G.~Guo, X.~-G.~He and G.~-N.~Li,
JHEP \textbf{1210}, 044 (2012) [arXiv:1207.6308 [hep-ph]];
M.~Aoki, J.~Kubo and H.~Takano,
Phys.\ Rev.\ D \textbf{87}, 116001 (2013) [arXiv:1302.3936 [hep-ph]];
Y.~Kajiyama, H.~Okada and K.~Yagyu,
Nucl.\ Phys.\ B \textbf{874}, 198 (2013) [arXiv:1303.3463 [hep-ph]];
Y.~Kajiyama, H.~Okada and T.~Toma,
Phys.\ Rev.\ D \textbf{88}, 015029 (2013) [arXiv:1303.7356];
  S.~Baek, H.~Okada and T.~Toma,
JCAP {\bf 1406}, 027 (2014)
[arXiv:1312.3761 [hep-ph]].
S.~Baek, H.~Okada and T.~Toma,
Phys.\ Lett.\ B {\bf 732}, 85 (2014)
[arXiv:1401.6921 [hep-ph]].
H.~Okada,
arXiv:1404.0280 [hep-ph];
M.~Aoki and T.~Toma,
JCAP \textbf{1409}, 016 (2014) [arXiv:1405.5870 [hep-ph]];
H.~Okada, T.~Toma and K.~Yagyu, 
Phys.\ Rev.\ D \textbf{90}, 095005 (2014) [arXiv:1408.0961 [hep-ph]];
  S.~Kashiwase, H.~Okada, Y.~Orikasa and T.~Toma,
Int.\ J.\ Mod.\ Phys.\ A {\bf 31}, no. 20n21, 1650121 (2016)
[arXiv:1505.04665 [hep-ph]].
  H.~Okada and Y.~Orikasa,
Phys.\ Rev.\ D {\bf 93}, 013008 (2016)
[arXiv:1509.04068 [hep-ph]];
  S.~Kanemura, K.~Nishiwaki, H.~Okada, Y.~Orikasa, S.~C.~Park and R.~Watanabe,
PTEP {\bf 2016}, no. 12, 123B04 (2016)
[arXiv:1512.09048 [hep-ph]].
S.~Kanemura, T.~Nabeshima and H.~Sugiyama,
Phys.\ Rev.\ D \textbf{85}, 033004 (2012) [arXiv:1111.0599 [hep-ph]];
  R.~Ding, Z.~L.~Han, Y.~Liao and W.~P.~Xie,
  JHEP {\bf 1605}, 030 (2016)
  doi:10.1007/JHEP05(2016)030
  [arXiv:1601.06355 [hep-ph]].
  Q.~H.~Cao, S.~L.~Chen, E.~Ma, B.~Yan and D.~M.~Zhang,
  Phys.\ Lett.\ B {\bf 779}, 430 (2018)
  [arXiv:1707.05896 [hep-ph]].
  R.~Ding, Z.~L.~Han, L.~Huang and Y.~Liao,
  Chin.\ Phys.\ C {\bf 42}, 103101 (2018)
  [arXiv:1802.05248 [hep-ph]].

\bibitem{three loops}M.~Aoki, S.~Kanemura and O.~Seto,
Phys.\ Rev.\ Lett.\ \textbf{102}, 051805 (2009) [arXiv:0807.0361 [hep-ph]];
M.~Aoki, S.~Kanemura and O.~Seto,
Phys.\ Rev.\ D \textbf{80}, 033007 (2009) [arXiv:0904.3829 [hep-ph]];
M.~Gustafsson, J.~M.~No and M.~A.~Rivera,
Phys.\ Rev.\ Lett.\ \textbf{110}, 211802 (2013) [arXiv:1212.4806
[hep-ph]]; 
A.~Ahriche and S.~Nasri,
JCAP \textbf{1307}, 035 (2013) [arXiv:1304.2055];
A.~Ahriche, K.~L.~McDonald and S.~Nasri,
JHEP \textbf{1410}, 167 (2014) [arXiv:1404.5917 [hep-ph]];
C.~S.~Chen, K.~L.~McDonald and S.~Nasri,
Phys.\ Lett.\ B \textbf{734}, 388 (2014) [arXiv:1404.6033 [hep-ph]].
H.~Okada and Y.~Orikasa,
Phys.\ Rev.\ D \textbf{90}, 075023 (2014) [arXiv:1407.2543 [hep-ph]];
  H.~Hatanaka, K.~Nishiwaki, H.~Okada and Y.~Orikasa,
Nucl.\ Phys.\ B {\bf 894}, 268 (2015)
[arXiv:1412.8664 [hep-ph]].
L.~G.~Jin, R.~Tang and F.~Zhang,
Phys.\ Lett.\ B \textbf{741}, 163 (2015) [arXiv:1501.02020 [hep-ph]];
P.~Culjak, K.~Kumericki and I.~Picek,
Phys.\ Lett.\ B \textbf{744}, 237 (2015) [arXiv:1502.07887 [hep-ph]];
A.~Ahriche, K.~L.~McDonald and S.~Nasri,
Phys.\ Rev.\ D {\bf 92} (2015) 9, 095020 [arXiv:1508.05881
[hep-ph]];
K.~Nishiwaki, H.~Okada and Y.~Orikasa,
Phys.\ Rev.\ D {\bf 92}, 093013 (2015)
[arXiv:1507.02412 [hep-ph]];

\bibitem{Law:2013saa}
  S.~S.~C.~Law and K.~L.~McDonald,
  JHEP {\bf 1309}, 092 (2013)
  [arXiv:1305.6467 [hep-ph]].
	
\bibitem{Ahriche:2016cio}
  A.~Ahriche, K.~L.~McDonald and S.~Nasri,
  JHEP {\bf 1606}, 182 (2016)
  [arXiv:1604.05569 [hep-ph]].

\bibitem{SInu}
R.~Foot, A.~Kobakhidze, K.~L.~McDonald and R.~R.~Volkas,
Phys.\ Rev.\ D \textbf{76} 075014 (2007) [arXiv:0706.1829 [hep-ph]];
S.~Iso, N.~Okada and Y.~Orikasa,
Phys.\ Lett.\ B \textbf{676}, 81 (2009) [arXiv:0902.4050 [hep-ph]];
  Z.~Kang,
  Eur.\ Phys.\ J.\ C {\bf 75}, 471 (2015)
  [arXiv:1411.2773 [hep-ph]];
  H.~Okada and Y.~Orikasa,
  Phys.\ Lett.\ B {\bf 760}, 558 (2016)
  doi:10.1016/j.physletb.2016.07.039
  [arXiv:1412.3616 [hep-ph]].
  J.~Guo, Z.~Kang, P.~Ko and Y.~Orikasa,
  Phys.\ Rev.\ D {\bf 91}, 115017 (2015)
  [arXiv:1502.00508 [hep-ph]];
  P.~Humbert, M.~Lindner and J.~Smirnov,
  JHEP {\bf 1506}, 035 (2015)
  [arXiv:1503.03066 [hep-ph]];
  P.~Humbert, M.~Lindner, S.~Patra and J.~Smirnov,
  JHEP {\bf 1509}, 064 (2015)
  [arXiv:1505.07453 [hep-ph]];
  A.~Ahriche, K.~L.~McDonald and S.~Nasri,
  JHEP {\bf 1602}, 038 (2016)
  [arXiv:1508.02607 [hep-ph]]; 
  H.~Okada, Y.~Orikasa and K.~Yagyu,
  arXiv:1510.00799 [hep-ph].
  V.~Brdar, Y.~Emonds, A.~J.~Helmboldt and M.~Lindner,
  arXiv:1807.11490 [hep-ph].

\bibitem{SIDM}
  J.~Guo and Z.~Kang,
  Nucl.\ Phys.\ B {\bf 898}, 415 (2015)
  doi:10.1016/j.nuclphysb.2015.07.014
  [arXiv:1401.5609 [hep-ph]].
  K.~Ghorbani and H.~Ghorbani,
  JHEP {\bf 1604}, 024 (2016)
  [arXiv:1511.08432 [hep-ph]].
  A.~D.~Plascencia,
  JHEP {\bf 1509}, 026 (2015)
  doi:10.1007/JHEP09(2015)026
  [arXiv:1507.04996 [hep-ph]].
  A.~Karam and K.~Tamvakis,
  Phys.\ Rev.\ D {\bf 94}, 055004 (2016)
  doi:10.1103/PhysRevD.94.055004
  [arXiv:1607.01001 [hep-ph]].
  P.~H.~Ghorbani,
  arXiv:1711.11541 [hep-ph].

\bibitem{SInuDM}
  H.~Davoudiasl and I.~M.~Lewis,
  Phys.\ Rev.\ D {\bf 90}, 033003 (2014)
  doi:10.1103/PhysRevD.90.033003
  [arXiv:1404.6260 [hep-ph]].
  A.~Karam and K.~Tamvakis,
  Phys.\ Rev.\ D {\bf 92}, 075010 (2015)
  doi:10.1103/PhysRevD.92.075010
  [arXiv:1508.03031 [hep-ph]].

\bibitem{Ahriche:2016ixu}
  A.~Ahriche, A.~Manning, K.~L.~McDonald, and S.~Nasri,
  Phys.\ Rev.\ D {\bf 94}, 053005 (2016)
  [arXiv:1604.05995 [hep-ph]].

	\bibitem{Gildener:1976ih}
  	E.~Gildener and S.~Weinberg,
  	Phys.\ Rev.\ D {\bf 13}, 3333 (1976).
  	doi:10.1103/PhysRevD.13.3333
	
\bibitem{TheMEG:2016wtm}
  A.~M.~Baldini {\it et al.} [MEG Collaboration],
  Eur.\ Phys.\ J.\ C {\bf 76}, 434 (2016)
  doi:10.1140/epjc/s10052-016-4271-x
  [arXiv:1605.05081 [hep-ex]].

  \bibitem{PDG18}
  M. Tanabashi {\it et al.} (Particle Data Group), Phys.\ Rev.\ D {\bf 98}, 030001 (2018).

\bibitem{Bertl:2006up}
  W.~H.~Bertl {\it et al.} [SINDRUM II Collaboration],
  Eur.\ Phys.\ J.\ C {\bf 47}, 337 (2006).
  doi:10.1140/epjc/s2006-02582-x

\bibitem{Araki:2011hm}
  T.~Araki, C.~Q.~Geng and K.~I.~Nagao,
  Phys.\ Rev.\ D {\bf 83}, 075014 (2011)
  doi:10.1103/PhysRevD.83.075014
  [arXiv:1102.4906 [hep-ph]].

\bibitem{Araki:2010zz}
  T.~Araki, C.~Q.~Geng and K.~I.~Nagao,
  Int.\ J.\ Mod.\ Phys.\ D {\bf 20}, 1433 (2011)
  doi:10.1142/S021827181101961X
  [arXiv:1108.2753 [hep-ph]].
	
\bibitem{Belanger:2018mqt}
  G.~Belanger, F.~Boudjema, A.~Goudelis, A.~Pukhov and B.~Zaldivar,
  Comput.\ Phys.\ Commun.\  {\bf 231}, 173 (2018)
  doi:10.1016/j.cpc.2018.04.027
  [arXiv:1801.03509 [hep-ph]].
	
\bibitem{Cui:2017nnn}
  X.~Cui {\it et al.} [PandaX-II Colla.],
  Phys.\ Rev.\ Lett.\  {\bf 119}, 181302 (2017)
  doi:10.1103/PhysRevLett.119.181302
  [arXiv:1708.06917 [astro-ph.CO]].
	
\bibitem{Aprile:2018dbl}
  E.~Aprile {\it et al.} [XENON Collaboration],
  arXiv:1805.12562 [astro-ph.CO].

\bibitem{Khachatryan:2016whc}
  V.~Khachatryan {\it et al.} [CMS Collaboration],
  JHEP {\bf 1702}, 135 (2017)
  doi:10.1007/JHEP02(2017)135
  [arXiv:1610.09218 [hep-ex]].

\bibitem{Aad:2015txa}
  G.~Aad {\it et al.} [ATLAS Collaboration],
  JHEP {\bf 1601}, 172 (2016)
  doi:10.1007/JHEP01(2016)172
  [arXiv:1508.07869 [hep-ex]].
	
	\bibitem{Farzinnia:2013pga}
  	A.~Farzinnia, H.~J.~He and J.~Ren,
  	Phys.\ Lett.\ B {\bf 727}, 141 (2013)
  	doi:10.1016/j.physletb.2013.09.060
  	[arXiv:1308.0295 [hep-ph]].

   \bibitem{Christensen:2008py}
 N.~D.~Christensen and C.~Duhr,
 Comput.\ Phys.\ Commun.\  {\bf 180}, 1614 (2009)
 [arXiv:0806.4194 [hep-ph]].
 N.~D.~Christensen, P.~de Aquino, C.~Degrande, C.~Duhr, B.~Fuks, M.~Herquet, F.~Maltoni and S.~Schumann,
 Eur.\ Phys.\ J.\ C {\bf 71}, 1541 (2011)
 [arXiv:0906.2474 [hep-ph]].
 A.~Alloul, N.~D.~Christensen, C.~Degrande, C.~Duhr and B.~Fuks,
 Comput.\ Phys.\ Commun.\  {\bf 185}, 2250 (2014)
 [arXiv:1310.1921 [hep-ph]].

 \bibitem{Degrande:2011ua}
 C.~Degrande, C.~Duhr, B.~Fuks, D.~Grellscheid, O.~Mattelaer and T.~Reiter,
 Comput.\ Phys.\ Commun.\  {\bf 183}, 1201 (2012)
 [arXiv:1108.2040 [hep-ph]].
 \bibitem{Alwall:2011uj}
 J.~Alwall, M.~Herquet, F.~Maltoni, O.~Mattelaer and T.~Stelzer,
 JHEP {\bf 1106}, 128 (2011)
 [arXiv:1106.0522 [hep-ph]].
 J.~Alwall {\it et al.},
 JHEP {\bf 1407}, 079 (2014)
 [arXiv:1405.0301 [hep-ph]].

 \bibitem{Ball:2012cx}
 R.~D.~Ball {\it et al.},
 Nucl.\ Phys.\ B {\bf 867}, 244 (2013)
 [arXiv:1207.1303 [hep-ph]].
 R.~D.~Ball {\it et al.} [NNPDF Collaboration],
 JHEP {\bf 1504}, 040 (2015)
 [arXiv:1410.8849 [hep-ph]].

 \bibitem{Sjostrand:2006za}
 T.~Sjostrand, S.~Mrenna and P.~Z.~Skands,
 JHEP {\bf 0605}, 026 (2006)
 [hep-ph/0603175].

 \bibitem{Ovyn:2009tx}
 S.~Ovyn, X.~Rouby and V.~Lemaitre,
 arXiv:0903.2225 [hep-ph].
 J.~de Favereau {\it et al.} [DELPHES 3 Collaboration],
 JHEP {\bf 1402}, 057 (2014)
 [arXiv:1307.6346 [hep-ex]].

 \bibitem{Conte:2012fm}
 E.~Conte, B.~Fuks and G.~Serret,
 Comput.\ Phys.\ Commun.\  {\bf 184}, 222 (2013)
 [arXiv:1206.1599 [hep-ph]].
 E.~Conte, B.~Dumont, B.~Fuks and C.~Wymant,
 Eur.\ Phys.\ J.\ C {\bf 74}, 3103 (2014)
 [arXiv:1405.3982 [hep-ph]].
 B.~Dumont {\it et al.},
 Eur.\ Phys.\ J.\ C {\bf 75}, 56 (2015)
 [arXiv:1407.3278 [hep-ph]].

 \bibitem{Drees:2013wra}
 M.~Drees, H.~Dreiner, D.~Schmeier, J.~Tattersall and J.~S.~Kim,
 Comput.\ Phys.\ Commun.\  {\bf 187}, 227 (2015)
 [arXiv:1312.2591 [hep-ph]].
 D.~Dercks, N.~Desai, J.~S.~Kim, K.~Rolbiecki, J.~Tattersall and T.~Weber,
 arXiv:1611.09856 [hep-ph].

\bibitem{Aaboud:2018jiw}
  M.~Aaboud {\it et al.} [ATLAS Collaboration],
  arXiv:1803.02762 [hep-ex].

\bibitem{Aaboud:2017qph}
  M.~Aaboud {\it et al.} [ATLAS Collaboration],
  Eur.\ Phys.\ J.\ C {\bf 78}, 199 (2018)
  [arXiv:1710.09748 [hep-ex]].

\end{thebibliography}
\end{document}